\documentclass{article}
\usepackage[T1]{fontenc} 
\usepackage[utf8]{inputenc} 
\usepackage{ismir,amsmath,cite,url}
\usepackage{graphicx}
\usepackage{color}
\usepackage[bookmarks=false]{hyperref}
\usepackage{tabularx,ragged2e,booktabs,caption}
\usepackage{array}
\usepackage{subcaption}
\usepackage{lineno}

\title{Towards a computational definition of the Tresillo rhythm and its tracing in Popular Music}

\threeauthors
  {Pushkar Jajoria} {EPFL$^{\dagger}$ \& USI}
  {Florian Krenn} {EPFL}
  {Aurel M\"ader}{EPFL}

\begin{document}

\maketitle

\begin{abstract}
This paper discusses the use and popularity of a rhythm, which   henceforth is referred to as ``Tresillo rhythm''. Thus we  first define and formalizes the Tresillo rhythm. Given a mathematical representation of the rhythm, the rhythm  is then traced in the US Billboard Top 20 Charts of the last 20 years. To detect and determine the use of the Tresillo rhythm in a song, similarities are calculated between a given formalization of the rhythm and  a given song. The calculated similarity, then indicate how similar the rhythm of a pop song is compared to the prior defined Tresillo rhythm.
To assert and  cross-validate  the computed rhythm similarity, two different formalizations of the Tresillo rhythm have been compiled and several different approaches to calculated rhythm similarities have been tested and compared. This similarity measure is then used to do an empirical study on the usage of the Tresillo rhythm in the US Billboard Top 20 Charts of the past 20 years (1999-2019). Finally, we argue about some of the possible reasons for the observed trend.\par 


\end{abstract}

\section{Research Question}

Can it be computed to which extent the Tresillo rhythm is used in a given pop song and if so how has the intensity of Tresillo rhythm use in the US Billboard Top 20 Charts changed over time?

\section{Introduction} \label{sec:introduction} 
The Tresillo is a rhythm that originated in Africa and was brought to the Caribbeans during the Atlantic Slave Trade period. Made popular in Cuba, the rhythm spread all over the world from there\cite{acquista2009tresillo, floyd1999black}, and can be found in many music genres. \par
\begin{figure}
 \includegraphics[width=\columnwidth]{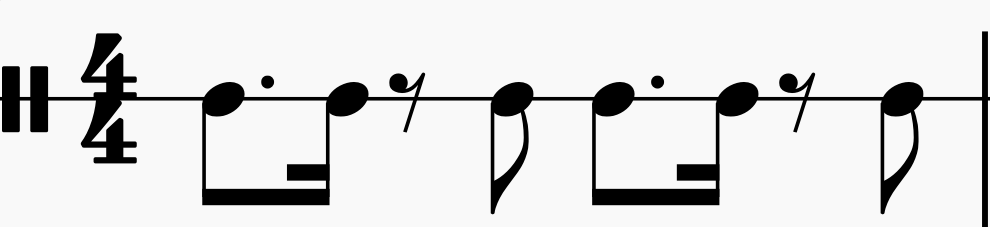}
 \caption{Synthetic Tresillo}
 \label{fig:tresillo}

 \includegraphics[width=\columnwidth]{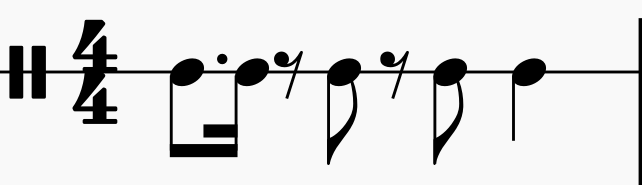}
 \caption{Clave rhythm}
 \label{fig:clave}

 \includegraphics[width=\columnwidth]{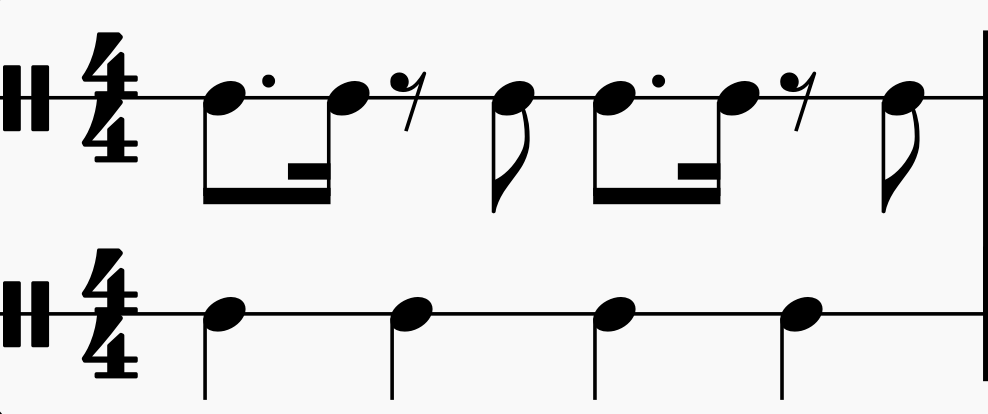}
 \caption{Reggaeton Rhythm}
 \label{fig:reggaeton}
\end{figure}

While being used as main rhythm on its own, the Tresillo is also used as rhythmic pattern in other rhythms such as the Reggaeton rhythm or the Clave rhythm.
Orientated on \cite {floyd1999black}, the Tresillo rhythm can be defined as followed (see Figure \ref{fig:tresillo}).

The rhythm pattern consists of a dotted eighth note, followed by a sixteenth note, an eighth rest and an eighth note and is repeated two times in a 4/4 bar.
If one adds a beat on all fours to the rhythm, one obtains the  Reggaeton rhythm (see Figure \ref{fig:reggaeton}). By changing the second half of the bar to a eighth rest, two eighth notes and another eighth rest, one creates the basic Clave rhythm pattern (see Figure \ref{fig:clave}).

The Tresillo rhythm is commonly used in latin american music.
However, the use of this  ``danceable Cuban Clave son"\cite{sethares2014geometry} is not restricted to Latin American music only, but also entered the rhythm sections of  Western  music \cite{sethares2014geometry, biamonte2018rhythmic}.
Popular recent examples include songs like ``Shape of you" by Ed Sheeran or ``Cheap Thrills" by Sia, both topping the Billboard Charts. Further investigation of the pop charts by the authors by listening to the Billboard Charts confirms the regular use of this rhythm in Pop Music. While the use, evolution and popularity of the Tresillo rhythm has been explored in qualitative  studies \cite{acquista2009tresillo, sandroni2001tresillo, sethares2014geometry, biamonte2018rhythmic, floyd1999black}, the popularity of this rhythm has not been studied quantitatively before.
In search of a possibility to investigate the use of this specific rhythm, we developed a method to computationally represent the main rhythm of a pop song and the Tresillo rhythm and compare both with different similarity measurements.  Given these similarity measures, a distinct time trend in Tresillo use can be found in the US Billboard Top 20 Charts of the last 20 years (1999-2019). \par
This paper proceeds as follows: First, the secondary literature of several different fields, which are relevant to this paper, will be discussed. Then several assumptions necessary to conduct the presented analysis will be stated in the problem statement section. The Data section discusses the chosen data sources and data format for the analysis. In the method section, the final data representation, the proposed rhythm similarity measures and evaluation metrics are presented and explained. In the results section, the different rhythm similarity measures are evaluated and compared. The results section also comprises a description and analysis of the time trend. The paper concludes with a  discussion of the chosen methods and obtained results. Furthermore, a possible interpretation of the produced results is presented. Lastly, we suggest possible extensions of the presented work.

\section{Secondary Literature}
This paper touches upon different scientific fields such as  musicology, audio retrieval and digital musicology. \par
Prior works have already investigated the evolution and spread of the Tresillo and Clave rhythm patterns and thus provide a clear definition and formalization of those rhythmic patterns from a theoretical view point \cite{acquista2009tresillo, sandroni2001tresillo, floyd1999black}. We heavily rely on those theoretical accounts to define the Tresillo rhythm used in this project.  Music scholar furthermore  investigated the diffusion of the Tresillo rhythm from Africa to  Latin America and then to United States from a cultural perspective \cite{acquista2009tresillo, floyd1999black}. More generally, there also have been several works discussing the rise in popularity of  Latin American music and its influence  onto U.S. mainstream music \cite{roberts1999latin, schroeder1978growth}. However, the mentioned musicology research is predicated upon qualitative analyses of musicology books, sheet music, recordings and interviews with specialists and practitioners. This paper in contrast chooses to employ computational methods to draw conclusion about the influence and popularity of the Tresillo rhythm in US popular music. \par 
Another research area that is deeply connected to the discussed topics in this paper is concerned with the formalization of rhythm and the statistical corpus studies of rhythmic patterns. While theoretical formulations of rhythm \cite{london2012hearing, rohrmeier2020towards} help us to assess the posed problem and possible pitfalls of the chosen methodology, this paper mainly refers to rhythm representations, which were used for corpus studies \cite{huron2006empirical} or more generally the study of onset frequency distributions \cite{palmer1990mental}. To represent rhythm in this project we will thus employ rhythm histograms as used and described in prior works \cite{palmer1990mental, huron2006empirical}. \par
A last field of research that is highly relevant for this paper  is concerned with computing rhythmic similarity between different songs. Such techniques are often used for audio retrieval tasks \cite{foote2002audio} or music genre classification tasks \cite{dixon2004towards, peeters2005rhythm}. More generally, this literature is concerned with measuring similarity and dissimilarity of audio signals or signals in general \cite{Zhang}. This literature provides valuable metrics and techniques to compare the rhythmical structure of two songs, however, is mainly based on using raw audio files to extract signal features and more specifically rhythm features \cite{foote2002audio, dixon2004towards, peeters2005rhythm, panteli2014modeling}. Thus the methods proposed in those papers, have been adjusted to work with our already discretized data representation.\par
This paper extends on the discussed secondary literature by using computational methods to trace the usage of a specific rhythm, which is associated to Latin American musical culture, in US popular music.

\section{Problem statement}
To answer  the research questions, a way of defining the main rhythm of a pop song is necessary. The vast majority of pop songs consist of a simple melodic and rhythmic structure. We, therefore, assume that one can  identify one dominant rhythm per song. This rhythm is repeatedly played throughout the song and therefore can be characterised by counting the onsets and comparing the onsets counts for every bar position.  
 To present the music in a usable format, quantification is needed. 16th notes are chosen as the smallest unit. Assuming that all songs used for our analysis are in a 4/4 meter, this gives 16 possible events per bar. All songs in the data which are found not to be in 4/4 are excluded from the analysis. Aggregating all bar onsets of a song to one bar results in one bar which can be described as a 16-dimensional vector, where every value represents the number of onsets on a given bar position.
 The Tresillo rhythm is used as a rhythm on its own or as part of other more complex rhythms. For our definition of clean Tresillo rhythm, we use the notation in Figure \ref{fig:tresillo}.

\section{Data}
To answer the posed research question four different kinds of data sets from different sources are needed. \par
First, to evaluate the proposed methodology which aims to compute a similarity between a defined Tresillo rhythm and a given song, two validation data sets have been collected. 
Of which the first data set consists of Tresillo songs and the second of songs that do not contain the Tresillo rhythm. Both data sets have been evaluated and hand-selected by the authors them self, by listening to spotify songs and choosing suitable examples. 
More specifically, to obtain songs which contain the Tresillo rhythm, a pre-compiled spotify play list was evaluated, which claimed to contain Tresillo songs\footnote{\text{https://open.spotify.com/playlist/17Na5AMmlLwY7OTsw6ovsS}}. 
After obtaining artist and song names of suitable validation set songs, appropriated MIDI files were searched on MIDIdb\footnote{https://www.mididb.com/} and downloaded.
\par
To trace the Tresillo rhythm in the popular music of the past 20 years a publicly available data set which contains the song names and artist names of the Hot 100 US Billboard Charts (1999-2019) was used\footnote{https://www.kaggle.com/danield2255/data-on-songs-from-billboard-19992019}. However to reduce the complexity of the data collection, it was decided to only use the US Billboard Top 20 Charts (1999-2019), which consists of in total of 1'447 songs. Given artist and song names, a web scrapper was coded and used to collect the available songs of the US Billboard Top 20 Charts (1999-2019) from the website MIDIdb\footnote{https://www.mididb.com/}. The final Billboard  data set on which the analysis was conducted on, consists of 444 distinct Billboard Top 20 weekly songs, which represent around 31\% of the US Billboard Top 20 Charts.
To assert the representativeness of the collected sample, the sample distribution was compared to the ground truth distribution of the US Billboard Top 20 Charts by evaluating t-test statistics of several features (e.g.: weeks on charts, peak position in charts, date of release). All t-tests indicate that the two distributions are not significantly different. \par
\begin{figure}
 \includegraphics[width=\columnwidth]{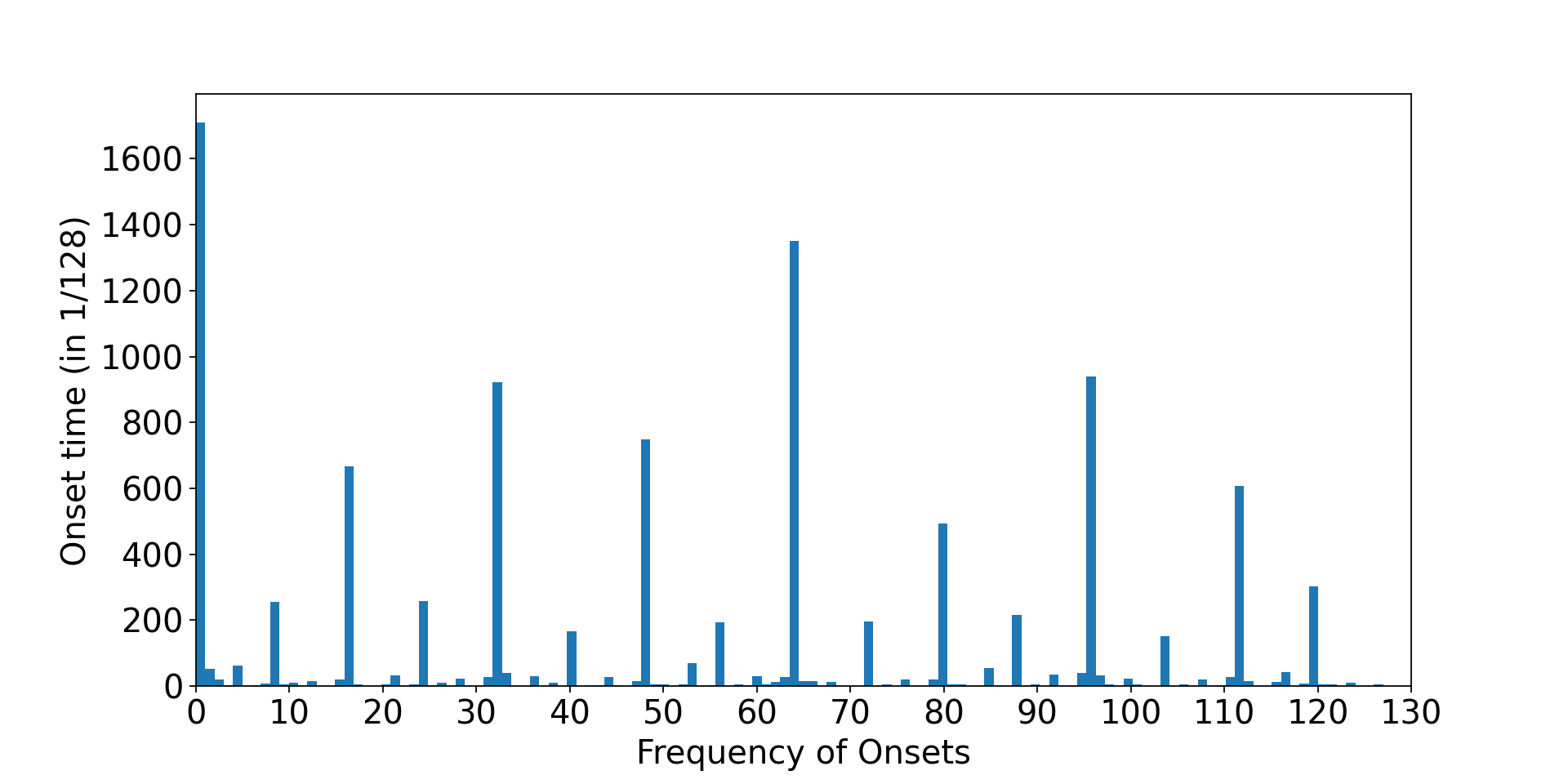}
 \caption{Frequency of onsets of musical events notated in 1/128 aggregated to one bar}
 \label{fig:onset_frequency_128}
\end{figure}

Initially, the data format of the collect musical data was MIDI. Audio was not used as initial data format, because obtaining onset tables for every voice would require complicated and elaborate computational processing of the data. MIDI has the practical advantage that, in contrast to other formats (e.g.: Musescore) it contains often multiple voices of a song. Furthermore, most pop songs are not available in score format.  However, to obtain onset lists for every musical event, the MIDI files have been converted to the Musescore format. Onset tables are then the final data representation used to obtain our results. With those onset tables Figure \ref{fig:onset_frequency_128} was compiled, which displays the frequency of onsets as notated in 1/128 notes aggregated to one bar.  \par
In addition, Muse Score provides the time signature for each song. The data set includes 9 songs in 3/4 or 6/8 time signature, which are excluded for further analysis.

\subsection{Defining the Tresillo rhythm}
To define the Tresillo rhythm computationally, the clean Tresillo rhythm as discussed in the introduction was used (see Figure \ref{fig:tresillo}). The clean version of the Tresillo is referred as Synthetic Tresillo in the following sections.

\section{Methods} 

\subsection{Rhythm vectors}
To be able to measure the similarity between two rhythms one must have a clear definition of rhythm. In general, one can define rhythm as "a series of onsets and durations of musical events.” \cite{rohrmeier2020towards}. Given that this paper investigates the dominant and repeating rhythm of a given song, it is however assumed that every musical event is sufficiently represented by its onset. \par
To obtain a computational representation of the dominant rhythm of a song, we aggregate all musical onset of a voice to one bar. Collapsing all musical onsets to one bar and thus obtaining onset 'histograms' is a common practice and has been used besides others to analyze Western  classical music \cite{palmer1990mental} and American folk music \cite{huron2006empirical}. \par


Given prior assessment of the Billboard data (see Figure \ref{fig:onset_frequency_128}) in conjunction with only working with songs with a 4/4 meter, this paper uses 16-dimensional vectors for the representation of rhythm. \par

This method provides for each voice of each song a 16 bin histogram denoting the cumulative number of onsets on a given beat. Here it is important to note that, although specific voices carry more information about the main rhythm of the song, considering each voice distinctly would require knowledge of which voice contributes how much to the perception of the main rhythm. As a method to obtain such knowledge is beyond the scope of this paper, no further steps are performed to differ between the voices.  By aggregating the onsets across all voices onto one single histogram, we obtain a single onset histogram for a given song. The onset histograms are then normalized to transform them into a 16 dimensional vector, which we will refer to as the rhythm vector of a song. This research assumes that the information about the rhythm of the song is captured by these rhythm vectors. \par
\begin{figure}
 \includegraphics[width=\columnwidth]{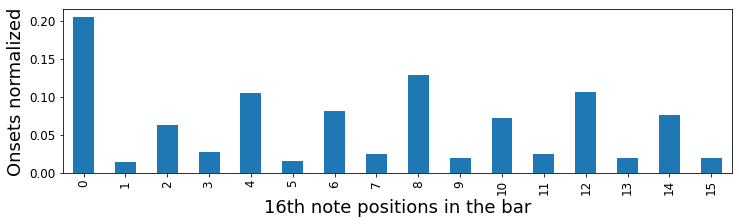}
 \caption{All rhythm vectors of all billboard songs aggregated to one vector}
 \label{fig:all_rhythmvectors}

 \includegraphics[width=\columnwidth]{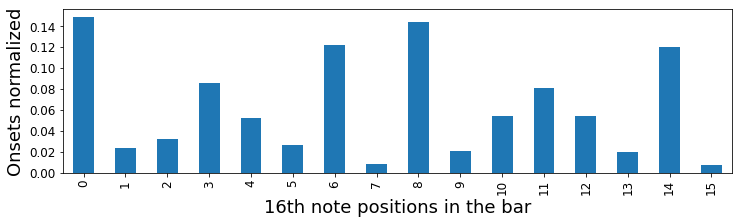}
 \caption{Rhythm vector of the song "Shape of you" by Ed Sheeran}
 \label{fig:shape of you vector}

 \includegraphics[width=\columnwidth]{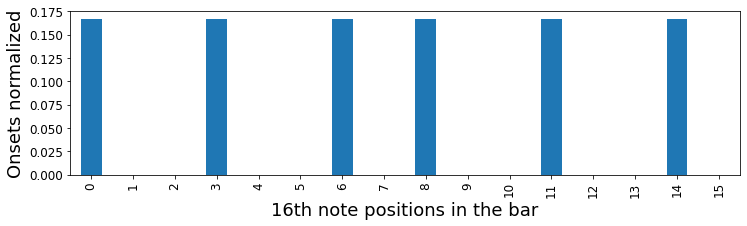}
 \caption{Rhythm vector of the synthetic tresillo}
 \label{fig:synthetic tresillo vector}
\end{figure}

The obtained rhythm vectors can be displayed as bar plots to allow visual inspection. Aggregating all rhythms vectors into one normalized rhythm vector shows the mean rhythm of our Billboard data set, as can be seen in Figure \ref{fig:all_rhythmvectors}. Figure \ref{fig:shape of you vector} shows a song with high Tresillo similarity and Figure \ref{fig:synthetic tresillo vector} the synthetic Tresillo pattern. Visual inspection and comparison of the compiled rhythm histograms, indicate similarities and differences between the rhythm vectors, which motivates our following methods. Thus, we present methods to systematically compare the rhythm vectors to each other in the following section. \par

\subsection{Tresillo similarity measures}
Each rhythm vector is a 16-dimensional vector for which similarity compared to another vector in the same space can be computed using a cosine similarity measure. We compare the similarity of rhythm vectors with two different Tresillo vectors which are defined as follows. 1) Template similarity center is the point in this vector space that corresponds to a plain Tresillo beat. 2) Centroid similarity point is defined as the centroid of all rhythm vectors corresponding to Tresillo songs.\par
The Tresillo rhythm is defined by its syncopated pattern. This information is visible in Figure \ref{fig:shape of you vector} by a sharp peak on the 3rd and 12th beat. Since each rhythm is defined by a higher and lower values along these 16 dimensions, it is fair to assume that each axis or onset position does not carry equal weight in the identification of this pattern. For example, it is very common in songs to have an onset at the start of a bar. Since the onset on the first beat is so ubiquitous in music, it will not carry a higher weight in the deduction of a rhythm in this vector space. To encapsulate this information into a similarity measure, we learn the scaling factors for each dimension of the rhythm space, which are referred by ${\theta}_i$. These ${\theta}_i$ are used to increase the gap between ``Tresillo similarities'' of rhythm vectors which do contain Tresillo and vectors which don't. These ${\theta}_i$ are scaling the rhythm vector of a song along that axis and in turn scale the similarity measure accordingly. The resulting parameterized cosine similarity is defined in equation \ref{eq_param_cosine} where, $\Theta$ refers to the set of scaling factors, A and B are two vectors with the same dimension between which the similarity needs to be computed. The $i^{th}$ dimension of these vectors are denoted by $a_i$ and $b_i$. $A_{\Theta}$ and $B_{\Theta}$ are the linearly transformed vector after scaling $i^{th}$ dimension by $\theta_i$. n denotes the total number of dimension in the rhythm space, i.e. 16 in our case.
Parameterized distance measures have been successfully used in the past in pattern recognition and machine learning\cite{Zhang}. \par

\begin{equation} \label{eq_param_cosine}
\begin{split}
\cos_{\Theta} ({\bf A},{\bf B})= \frac{{\bf A_{\Theta}}*{\bf B_{\Theta}}} {\|{\bf A_{\Theta}}\| \|{\bf B_{\Theta}}\|} \\
= \frac{ \sum_{i=1}^{n}{{\bf a}_i.{\theta_i}{\bf b}_i.{\theta_i}} }{ \sqrt{\sum_{i=1}^{n}{({\bf a}_i.{\theta_i})^2}} \sqrt{\sum_{i=1}^{n}{({\bf b}_i.{\theta_i})^2}} }
\end{split}
\end{equation}

Equation \ref{eq_param_cosine} defines the parameterized cosine similarity used in this research. The parameters for this model are learned by maximizing $S^*$ (defined in the next section), which can also be modeled into a minimization problem as formulated in Equation \ref{eq_argmin}.

\begin{equation} \label{eq_argmin}
\underset{\Theta}{\mathrm{argmin}} \frac{cos_{\Theta}(\mathcal{A'}, \mathcal{T})}{cos_{\Theta}(\mathcal{A}, \mathcal{T})}
\end{equation}

Where $\mathcal{A}$ is the set of songs with Tresillo present in them, $\mathcal{A'}$ is the set of songs with Tresillo not present in them and $\mathcal{T}$ refers to the reference point for computing the cosine similarity.

\subsection{Evaluation}
To evaluate the proposed similarity methods two different metrics were chosen to assess the variance produced by  a given model and to compare the model fits between different models. \par
To assess the variance in Tresillo similarity estimated by a given model, the bootstrapping method on the validation data sets was used. Thus using the Tresillo and the non-Tresillo validation data sets, we used a given model to calculated the mean Tresillo similarity in a given validation set and its 95\% confidence intervals, as obtained by bootstrapping.
The bootstrapping was performed with 1'000 draws with replacement. The number of samples per draw, correspond to the sample size of a given validation set (e.i.: either 9 or 10 samples).\par
To compare different models it was assumed that a good model would have high similarity for all songs which have a Tresillo pattern and a low similarity for songs which do not have such a pattern (see Equation \ref{equ:model_godness}).
This can be measured by defining 'Similarity Goodness' $S^*$ as the ratio of mean similarity in songs that have Tresillo and mean similarity of songs that do not. Higher value of $S^*$ denote high similarity for songs with Tresillo and low similarity for songs without Tresillo. Here similarity refers to the similarity computed between the rhythm vector of a song and a Tresillo rhythm vector.  This similarity could be computed by using either an unparameterized or a parameterized cosine similarity, depending on the different models defined in the previous section.
\begin{equation}
S^* = \frac{\text{mean similarity of songs with tresillo }}{\text{mean similarity of songs without tresillo}}
\label{equ:model_godness}
\end{equation}




\section{Results}

\subsection{Comparing the similarity measures}

\begin{figure}[ht]
 \includegraphics[width=\columnwidth]{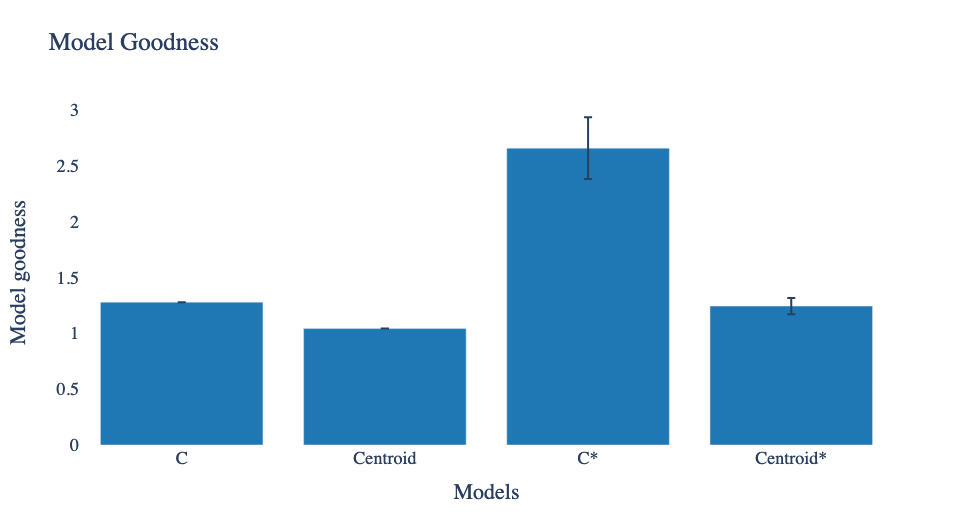}
 \caption{Comparing model goodness. Here, \textbf{C} refers to rhythm similarity measured with cosine similarity, using Tresillo template as centre.
 \textbf{Centroid} refers to rhythm similarity measured with cosine similarity, using the centroid of Tresillo songs as centre.
 \textbf{C*} refers to rhythm similarity measured with parameterized cosine similarity, using Tresillo template as centre.
 \textbf{Centroid*} refers to rhythm similarity measured with parameterized cosine similarity, using the centroid of Tresillo songs as centre.}
 \label{fig:model_goodness}
 
 \includegraphics[width=\columnwidth]{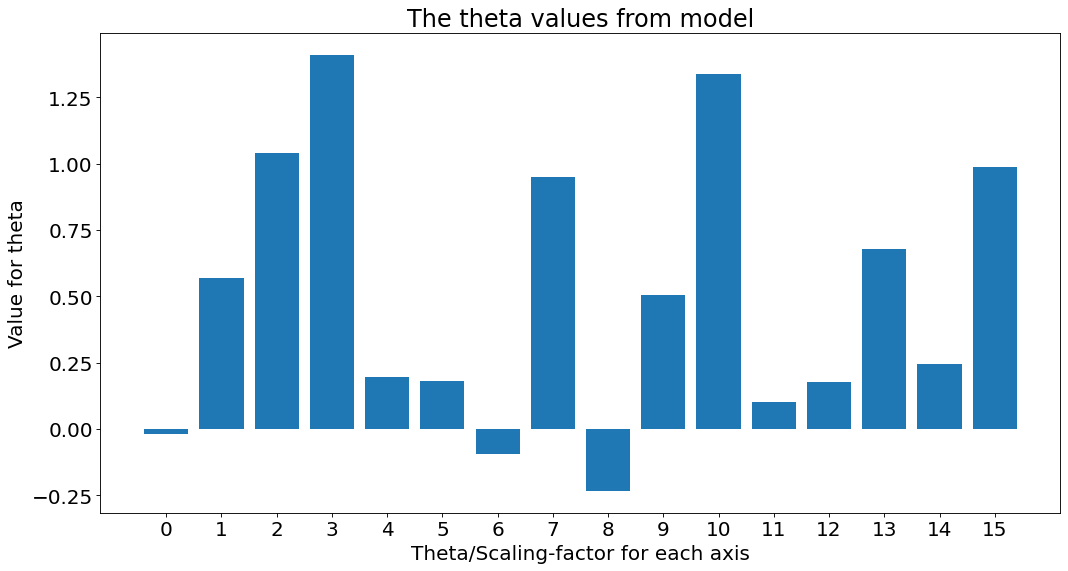}
 \caption{Theta for each dimension}
 \label{fig:model_parameters}
\end{figure}

Figure \ref{fig:model_goodness} compares $\mathcal{S*}$ using two different Tresillo centers with a parameterized vs unparameterized  cosine similarity. The error lines on the bar plot denote the 97.5\% confidence interval based on `leave one out` cross-validation.
It can be inferred that the models based on the synthetically defined Tresillo outperform the models based on the centroid methods(Both 'p' values < 0.001 using t-test). Parameterized models also outperform the un-parameterized models. (Both 'p' values < 0.001 using t-test) \par

Figure \ref{fig:model_parameters} shows the learned theta for each beat after fitting the model. Low and negative value for $0^{th} \text{ and } 8^{th}$ beat bolster our claim about the ubiquitous  $0^{th}$ beat in popular music. The third beat along with the second beat carry a lot of information about the Tresillo beat and hence has a high value. Other Tresillo beats also share a high peak with the exception of $6^{th}$ and the $8^{th}$ beat. This may be because the onsets for popular rock pop songs coincide here. There is asymmetry across the 8th beat. One of the possible reason for this could be the drum fills, which are often in the later half of the bar.
\\
\begin{table}
\centering

\begin{tabular}{ | m{10em} | m{1.5cm}| m{1.5cm} | } 

\hline
Song Name & $\mathcal{C}^*$ & $\mathcal{C}$\\
\hline
\hline
Cheap Thrills-Sia& 0.9919 & 0.9089 \\ 
\hline
Eastside-benny blanco & 0.9471 & 0.7112 \\ 
\hline
Let Me Love You-Justin Bieber & 0.2538 & 0.6051 \\ 
\hline
New rules-Dua Lipa& 0.9431 & 0.8388 \\ 
\hline
\end{tabular}
\caption{Cosine similarity and parameterized cosine similarity for popular Tresillo songs}{Here $\mathcal{C}^*$ Denotes parameterized cosine similarity and $\mathcal{C}$ denotes cosine similarity}
\label{tbl:similarity_table}

\end{table}

\subsection{Time Trend} 

Considering the evaluation of the proposed models based on our validation data sets, it was inferred that models using the synthetic defined Tresillo outperformed models using a data representation of the Tresillo. Thus to analyze the Tresillo time trend, only the two models, which are based on the synthetic Tresillo were used. Those two models correspond to the models with best performance on the validation set. However, it must be mentioned that the parametrized model clearly outperformed the cosine similarity model. \par
\begin{figure}
 \includegraphics[width=\columnwidth]{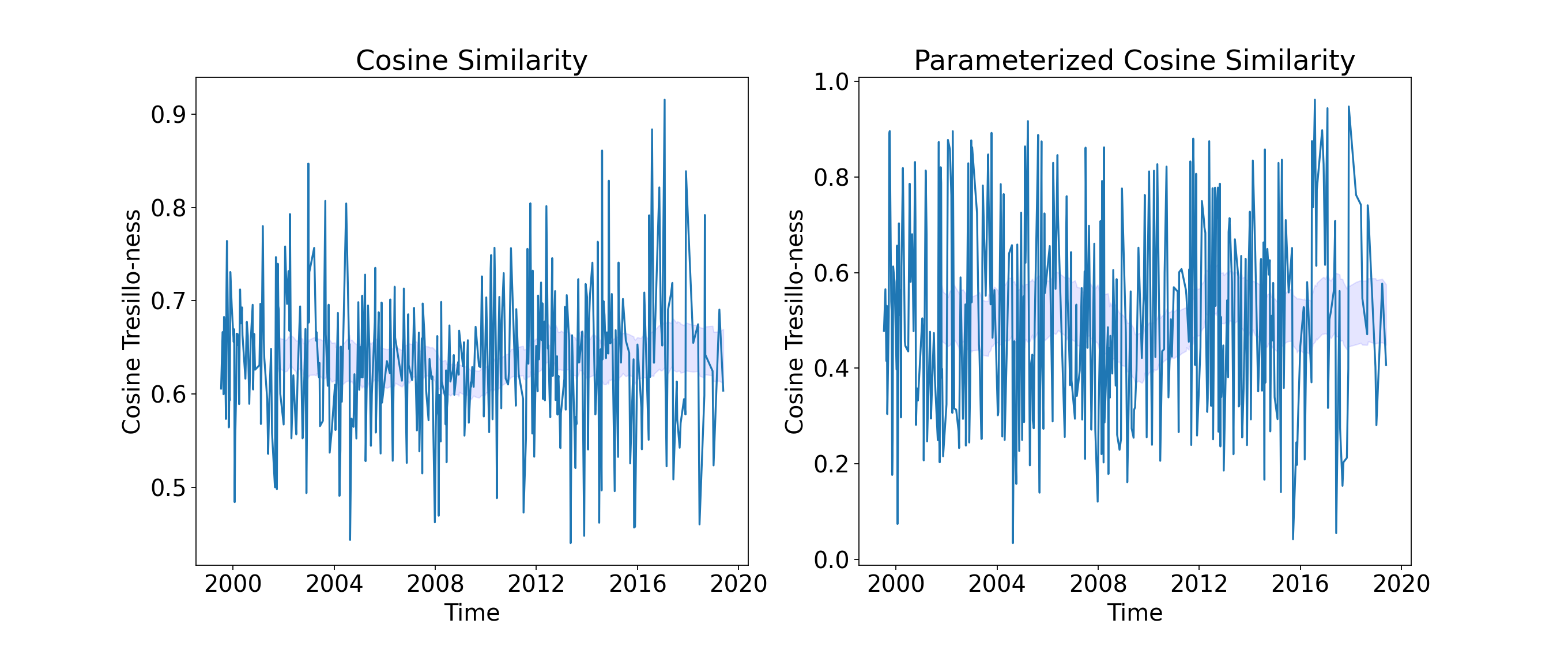}
 \caption{Weekly average of cosine Tresillo similarity in the US Billboard Top 20 Charts}
 \label{fig:tresilloness_rollin_week}
 \includegraphics[width=\columnwidth]{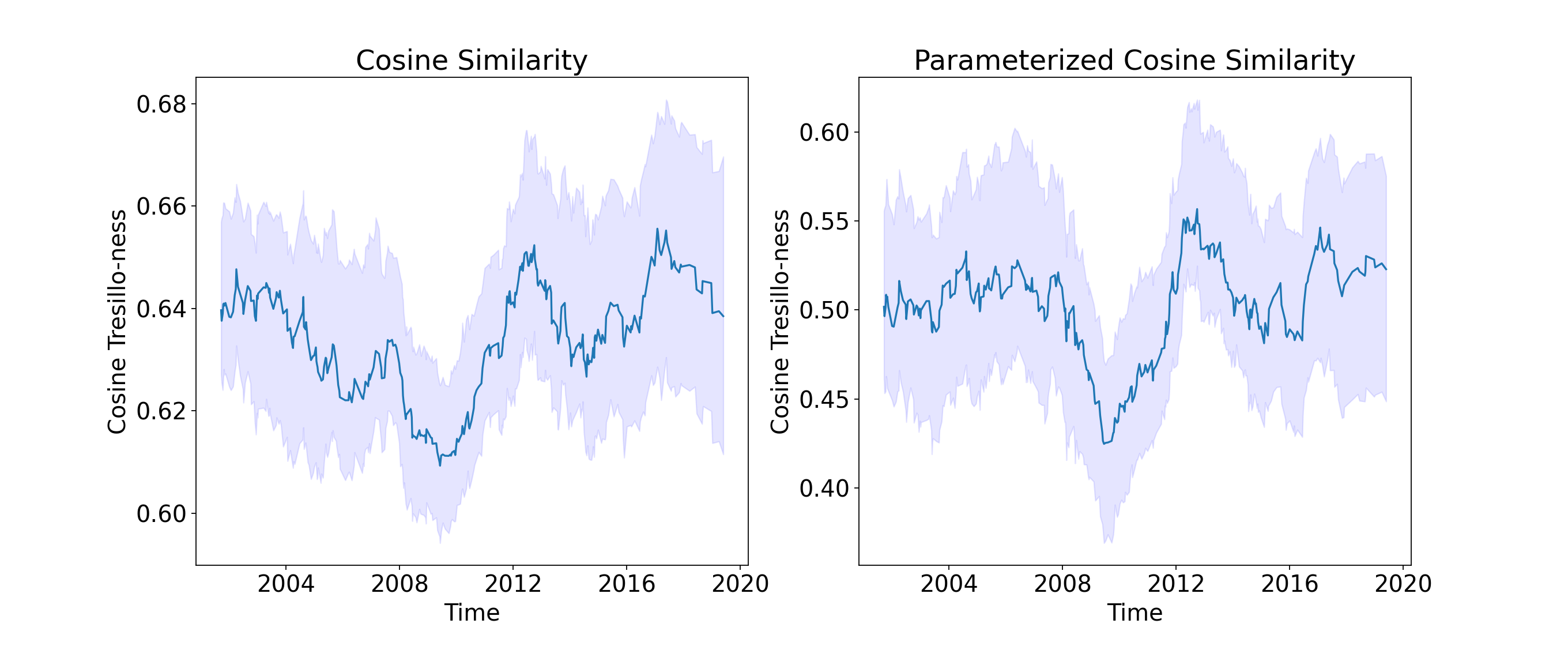}
 \caption{52 weeks moving average of cosine Tresillo similarity in the US Billboard Top 20 Charts}
 \label{fig:tresilloness_rolling_yearly}
\end{figure}

Given those two models, Tresillo similarity for all songs in the Billboard data set were calculated. By plotting the Tresillo similarity over time, a proxy for the trend in Tresillo use over time was obtained. In a first naive analysis, the mean weekly use of the Tresillo rhythm in the US Billboard Top 20 Charts can be seen in Figure \ref{fig:tresilloness_rollin_week}. To reduce the noise and variance visible in Figure \ref{fig:tresilloness_rollin_week}, a rolling yearly mean was applied to the weekly mean value in Tresillo use. The resulting rolling yearly average of Tresillo use can be seen in Figure  \ref{fig:tresilloness_rolling_yearly}. In both Figures, 95\% confidence intervals have been obtained via bootstrapping (1'000 draws with replacement, with the number of samples per draw equal to the sample size) and are indicated by a light blue coloring. 

\section{Discussion} 

This paper shows, that, given a clear definition, the intensity of  Tresillo rhythm use in a  given song can be measured with computational methods. Several methods have been introduced which  identify the use of the Tresillo rhythm and its intensity in the collected validation data sets. However, it is still disputable how the proposed models deal with noise which might dilute the Tresillo rhythm. Assessing the uncertainty of our models (e.g.: by looking at outliers as determined by the bootstrapping method), it is notable that not every song which was labeled to contain the Tresillo rhythm, has a very high Tresillo similarity. This is exemplified by Table \ref{tbl:similarity_table}. Although both model in Table \ref{tbl:similarity_table} have high similarity for a Tresillo song and low similarity for a 'non-Tresillo' song, parameterized model $\mathcal{C}^*$ has a larger gap between the two.\par
Assessing the time trends of Tresillo use in the US Billboard Top 20 Charts of the past 20 years, no clear linear time trend is observable. However, several interesting peaks and patterns are noticeable. First, looking at Figure \ref{fig:tresilloness_rollin_week} it is observable that the obtained results are by nature very noisy and there is high variance in Tresillo use from week to week. \par
Using a rolling yearly mean, the  trend of Tresillo use over time gets more visible as can be seen in Figure \ref{fig:tresilloness_rolling_yearly}. Figure \ref{fig:tresilloness_rolling_yearly} offers us some interesting insights. Even though there is considerable yearly variance in the intensity of Tresillo rhythm use (as illustrated by the 95\% confidence intervals), there are identifiable peaks and valleys in Tresillo rhythm use. Furthermore, the unparameteriaed cosine similarity and the parametrized cosine similarity produce consistent results, however on different scales. \par
Looking at Figure \ref{fig:tresilloness_rolling_yearly}, we observe a trend which starts relatively high around the new millennial, stays constant or slightly decreases till around 2008 from where on the trend collapses to its all time low in 2010. From 2010 on there is an increase in Tresillo use, although there is another valley around 2014. Finally around 2018 the trend in Tresillo rhythm intensity reaches its all time high. \par
After subjectively evaluating the calculated Tresillo similarity and the corresponding Billboard Charts, we interpret the prior described trend as follows. In the early 2000 it seems that many songs which peaked the Billboards were either by Latin artist or by artist, who used Latin music themes in their songs (e.g.: Maria Maria, Santana, 2000; Be With You, Enrique Iglesias 2000; Baby Boy, Beyonce, 2003). 
This trend then decreases steadily, till the Tresillo rhythm seems to  reappear in Western dance music after 2010. After 2010 there are several peaks which can be associated to popular dance music songs with particular high Tresillo similarity (e.g.: Where Have You Been, Rihanna, 2012; Shape Of You, Ed Sheeran, 2017; Cheap Thrills, Sia, 2016). However, to substantiate this interpretation of the time trend further empirical research would be needed, which clearly defines and differentiates the use of the Tresillo rhythm in the context of 'Latin American' music and its usage in Western dance music.

\section{Conclusion} 
This paper formalizes a mathematical representation of the Tresillo rhythm  and offers a methodology to compute the intensity of the Tresillo rhythm in a given song. It uses this methodology to trace the intensity of Tresillo use in the US Billboard Top 20 Charts (1999-2019). \par
This paper evaluates and compares several models to compute Tresillo similarity and tests the performance of the given models on validation data sets. Furthermore, the uncertainty of the obtained results is quantified. \par
Assessing the obtained time trend, distinct peaks and valleys can be observed, however there seem to be no linear time trend in the use of the Tresillo rhythm. After a relatively high starting level in Tresillo intensity around the new millennial, the average Tresillo similiarity decrease till 2010. Then there is a quadratic trend pointing towards increasing use of this rhythm.\par
By subjectively evaluating some Billboard charts and their corresponding Tresillo similarity, we interpret this trend to be explained by an initial high popularity of Latin music and after 2010 to the increasing use of this rhythm in Western dance music. However, further research would be needed to empirically substantiate this claim.

\section{Future Work}
The channels in a song carry valuable information and could be leveraged upon if a sophisticated algorithm could be developed which is agnostic to meta data information but rather works on a symbolic level. The above work assumes a 4/4 meter for a song, this assumption could also be removed by developing a algorithm to map songs with different time signature (3/4, 7/4) into the same rhythm space. \par
The lack of well annotated midi data is also a limiting factor. Annotating more data will result in better parameterized models which in turn, would improve the $\mathcal{S}^*$. The benefits of more data are not limited to this. More sophisticated learning algorithms which could not be used given over-fitting concerns, might become viable. For instance, a non linear transformation of the rhythm vector space may results in better results as this would be better suited at modeling the nuances on, for instance the 3rd beat. Dimension reduction techniques like PCA could also be employed to reduce over-fitting.\par
Finally, to substantiate our subjective impression that there are two waves in usage of this rhythm, once in Latin American music and once in Western dance music, further empirical research would be needed, which differentiates in which musical context this rhythm is used. 
\bibliography{main}

\end{document}